\documentclass[twocolumn,amssymb,floatfix,deluxe,prd,showkeys]{revtex4-1}
\usepackage{graphicx,%  Default Latex 2eps package for embedding figures
  longtable%              Useful for long table
}
\usepackage{dcolumn}% Align table columns on decimal point
\usepackage{bm}% bold math

\newcommand{\NotreDame}{University of Notre Dame, Notre Dame, IN 46556-5670, USA}
\newcommand{\Boston}{Boston University, Boston, Massachusetts 02215, USA}
\begin{document}
\title{{\em Comment on} Search for Muon Neutrino Oscillations with the
  Irvine-Michigan-Brookhaven Detector}
\author{J.M.~LoSecco}
\affiliation{\NotreDame}
\author{L.R.~Sulak}
\affiliation{\Boston}
\date{\today}
\begin{abstract}
  The neutrino flux used in Becker-Szendy {\em et al.}\cite{BZ} is now known to be
  in error.  Becker-Szendy {\em et al.}\cite{BZ} depended heavily on the
  Lee and Koh flux\cite{LeeKoh} which was noted as erroneous in Gaisser {\em et al.}\cite{Bug}.
  This makes the results of Becker-Szendy {\em et al.}\cite{BZ} unreliable.
\end{abstract}
\keywords{neutrino anomaly, neutrino oscillations, atmospheric neutrinos}
\maketitle
In an attempt to understand the atmospheric neutrino anomaly\cite{First} we
studied the higher energy upward going atmospheric muon neutrino flux by using
upward through-going and upward stopping muons created by neutrino interactions
in the rock surrounding the IMB detector\cite{BZ}.  Such an analysis depends
strongly on the neutrino flux estimate since neither the interaction vertex nor
the neutrino energy can be reconstructed.  The fiducial mass depends on the
neutrino energy and direction since higher energy muons will penetrate further
through the rock.  The work in \cite{BZ} depended on the calculations of
Lee and Koh\cite{LeeKoh} for the lower portion of the spectrum and
Volkova\cite{Volkova} for the high energy part.  This is risky since the overall
error in the normalization of the flux estimate can be reduced with relative
measurements of one portion of the flux to another portion.  This may not happen
when independent flux estimates are used for different energy regions.
It has come to our attention\cite{Olga} that \cite{BZ} disagrees, in part, with
current interpretations of the atmospheric neutrino anomaly as a result of
neutrino oscillations.

Gaisser {\em at al.}\cite{Bug} points out ``We have discovered several bugs in
the implementation of the LK code.'' in reference to \cite{LeeKoh} and drops it
from consideration. Gaisser {\em at al.}\cite{Bug} has an extensive discussion
of how errors in the neutrino flux calculations will lead to errors in
interpretation.

In light of the known error in \cite{LeeKoh} the results of \cite{BZ} are
unreliable.  A contemporary article \cite{Hirata} compared the low energy Lee
and Koh spectrum\cite{LeeKoh} with two other predictions, for both the Kamioka
and IMB sites.  The erroneous Lee and Koh spectrum consistently predicted lower
neutrino event rates.  Comparison of the Lee and Koh flux with modern estimates
indicates it was low by more than a factor of two at modest energies.  A later
measurement of the upward stopping muon fraction\cite{SuperK} gave an
observation consistent with Becker-Szendy {\em et al.}\cite{BZ}.  The more
reliable
flux estimates used in \cite{SuperK} predicted a stopping fraction which was
1.7$\pm$0.2 times higher than observed.

In spite of these normalization problems Lee and Koh gave a consistent result
of 1/2 for the electron to muon neutrino flux ratio.

We would like to thank Olga Botner for bringing the issue of \cite{BZ} to
our attention.  We would like to thank Bob Svoboda for help with the
comparisons.

\end{document}